\documentclass[fleqn,twoside]{article}
\usepackage{amsmath}
\usepackage{amssymb}
\usepackage[headings]{espcrc2}
\usepackage{psfrag}
\usepackage{slashed}

\usepackage{graphicx}

\def\fl#1{{#1}^\flat}
\def\kf#1{\fl{K_{#1}}}

\def\AB#1#2#3{\langle#1|#2|#3]}

\def\A#1#2{\langle#1#2\rangle}

\def\ra{\rangle}
\def\la{\langle}
\def\wh#1{\widehat{#1}}
\def\inf{{\rm Inf}}
\def\res{{\rm Res}}

\title{Generalised Unitarity At One-Loop With Massive Fermions}

\author{S. D. Badger\address{%
Institut de Physique Th\'{e}orique,
CEA-Saclay, F-91191 Gif-sur-Yvette, France\\
E-mail: {\tt simon.badger@cea.fr}}%
\thanks{I would like to thank David Kowoser for collaboration in the early stages of this work. I am
also grateful to Jurgen K\"orner and Zakaria Merebashvili for assistance with comparisons to their
Feynman computation. Supported by Agence Nationale de Recherche grant
ANR-05-BLAN-0073-01. 
}}

\begin{document}

\begin{abstract}
We describe an application of generalised unitarity to the computation of one-loop amplitudes with
massive external fermions. We present analytic results for the cut-constructible parts of the
leading colour contributions to the all-plus helicity configuration of the $t\bar{t}gg$ amplitude. Using a
special choice for the helicity basis of the massive fermions we are able to obtain extremely
compact analytic expressions. In particular we describe how one can fix the divergent contributions
from tadpole and wave-function renormalisation using universal UV and IR behaviour.
\vspace{1pc}
\end{abstract}

\maketitle

\section{Introduction}

The Large Hadron Collider, due to start later this year, will require accurate QCD predictions for
successful data analysis. New TeV scale physics is likely to
be associated with the production and subsequent decay of heavy particles and hence background
signatures will come in the form of multi-jet final states. As quantitative QCD predictions only
begin at NLO the main ingredient for the computation of such backgrounds are multi-particle one-loop
scattering amplitudes. Traditional Feynman methods struggle to cope with the enormous growth in the
number of diagrams with the number of external legs and so many processes remain uncomputed. As an
alternative, on-shell methods provide an extremely efficient tool by keeping only the
physical degrees of freedom, simplifying calculations. For a review of the subject we refer
the reader to reference \cite{Bern:review}.

Unitarity based techniques \cite{Bern:uni1,Bern:uni2} have been used successfully for many years to
compute complicated multi-particle and multi-loop gauge theory processes. Recent developments
utilising complex momenta have lead to new recursive techniques at both tree level \cite{Britto:rec}
and one-loop \cite{Bern:1lrec} as well as improved multiple cutting techniques for logarithmic terms
\cite{Britto:genu}.

A general one-loop amplitude can be written in terms of a basis of scalar box, triangle, bubble and tadpole functions together
with rational terms missed by standard four-dimensional cuts. Because the quadruple cut completely
freezes the loop integration, determination of the box coefficients is a completely algebraic
procedure. For the lower-point integral functions Ossola, Papadopoulos and Pittau have shown how to
systematically generate subtraction terms to find algebraic procedures for the remaining
coefficients \cite{Ossola:OPP}. By using a special complex parametrisation of the cut momenta, Forde has demonstrated that this method
can also be used to generate compact analytic expressions without subtraction \cite{Forde:intcoeffs}.
Initial numerical implementations, such as \texttt{BlackHat} \cite{Berger:blackhat} and
\texttt{Rocket} \cite{Giele:ddnumgu,Giele:rocket}, are making the first steps towards a NLO event generator.

In this paper we demonstrate the power and diversity of these new tools by computing the
cut-constructable contributions to top pair production via gluon fusion. We derive
compact analytic expressions in agreement with much lengthier expressions obtained from
previous Feynman diagram based computations \cite{Korner:ttgg}. This builds upon a growing body of work generalising unitarity
methods to massive processes \cite{Kilgore:massuni,Britto:gumassive,Britto:massive}.

Throughout this paper all amplitudes are considered to be colour-ordered helicity amplitudes, using
the standard spinor-helicity formalism to write the amplitudes in terms of spinor products. We accommodate massive momenta into
the formalism by decomposing them into two massless momenta using an arbitrary massless reference vector, $\eta$ \cite{Kleiss:massspin}:
\begin{equation}
	p = \fl p + \frac{m_p^2}{\AB{\eta}{p}{\eta}}\eta.
	\label{eq:massdecomp}
\end{equation}
This then allows the definition of external fermion wavefunctions as follows \cite{Schwinn:allborn,Rodrigo:heavyQ},
\begin{align}
	u_\pm(q,m) &= \frac{(\slashed{q}+m)|\eta\mp\ra}{\A{\fl q\pm|}{\eta\mp}}, \\
	v_\pm(q,m) &= \frac{(\slashed{q}-m)|\eta\mp\ra}{\A{\fl q\pm|}{\eta\mp}}.
\end{align}

\section{Integral Coefficients with massive propagators}

In this section we outline the method for direct extraction of integral coefficients with arbitrary
internal and external masses \cite{Forde:intcoeffs,Kilgore:massuni}. Throughout the next section we
take all external momenta to be out-going and all loop momenta to circulate in a clockwise
direction.

\subsection{Box Coefficients}

\begin{figure}[t]
	\begin{center}
		\psfrag{K1}{\tiny$K_4$}
		\psfrag{K2}{\tiny$K_1$}
		\psfrag{K3}{\tiny$K_2$}
		\psfrag{K4}{\tiny$K_3$}
		\psfrag{l}{\tiny$l_1$}
		\psfrag{l-K2}{\tiny$l_1-K_1$}
		\psfrag{l-K2-K3}{\tiny$l_1-K_1-K_2$}
		\psfrag{l+K1}{\tiny$l_1+K_4$}
		\includegraphics[width=0.5\columnwidth]{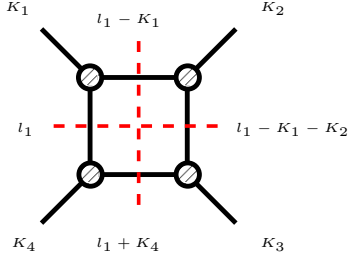}
	\end{center}
	\caption{Quadruple cut for box coefficients with all momenta out-going and loop momentum
	flowing clockwise.}
	\label{fig:box}
\end{figure}

For the coefficients of the box integrals, the on-shell constraints on the loop momenta freeze the
loop integration resulting in a purely algebraic procedure for the evaluation of the quadruple cut
\cite{Britto:genu}. Here we use a loop momentum parametrised by
\begin{align}
	l_1 = a \kf 4 + b \kf 1 + c|\kf 4\ra[\kf 1| + d|\kf 1\ra[\kf 4|,
\end{align}
where
\begin{align}
	&\kf 4 = \frac{\gamma K_4 - S_4K_1}{\gamma^2-S_1S_4}, \label{eq:fldefs1}\\
	&\kf 1 = \frac{\gamma K_1 - S_1K_4}{\gamma^2-S_1S_4}, \label{eq:fldefs2}\\
	&\gamma = K_1\cdot K_4\pm\sqrt{(K_1\cdot K_4)^2-S_1S_4}.\label{eq:fldefs3}
\end{align}
One can find the values of $a,b,c,d$ from the constraints,
\begin{align}
	\mathcal{S} = &\{l_1^2=m_1^2,(l_1-K_1)^2=m_2^2,\nonumber\\
	&(l_1-K_1-K_2)^2=m_3^2,(l_1+K_4)^2=m_4^2\};
\end{align}
explicitly this gives us
\begin{align}
	a &= \frac{S_1\wh{S}_4+\gamma\wh{S}_1}{\gamma^2-S_1S_4}, &
	b &= -\frac{S_1\wh{S}_4+\gamma\wh{S}_1}{\gamma^2-S_1S_4},\\
	d &= \frac{1}{c}\left( ab-\frac{\mu^2}{\gamma} \right),
\end{align}
and $c$ as a solution to the quadratic equation:
\begin{align}
	&c^2\AB{\kf 4}{K_2}{\kf 1}
	+  \left(ab-\frac{m_1^2}{\gamma}\right)\AB{\kf 1}{K_2}{\kf 4}\nonumber\\
	&+ c\left(a\AB{\kf 4}{K_2}{\kf 4}+b\AB{\kf 1}{K_2}{\kf 1}-\wh{S}_{12}\right)
	=0.
	\label{eq:ceq}
\end{align}
Here the mass dependence is determined by $\wh{S}_4=S_4+m_1^2-m_4^2$, $\wh{S}_1=S_1+m_1^2-m_2^2$ and
$\wh S_{12}=S_2+2K_1\cdot K_2+m_2^2-m_3^2$.
We label the complex solutions for the loop momentum as $l_1^\pm$, corresponding to the solutions
$c_\pm$ of eq. \eqref{eq:ceq}. The value of the box coefficient
is then given simply as the product of four tree
amplitudes evaluated at the values, $l_1^\pm$, of the loop momentum:
\begin{equation}
	C_{4;K_1|K_2|K_3}^{m_1m_1m_2m_3m_4} = \frac{i}{2}\sum_{\sigma=\pm} A_1A_2A_3A_4(l_1^\sigma).
	\label{eq:boxcoeff}
\end{equation}

\subsection{Triangle Coefficients}

\begin{figure}[t]
	\begin{center}
		\psfrag{K1}{\tiny$K_3$}
		\psfrag{K2}{\tiny$K_1$}
		\psfrag{K3}{\tiny$K_2$}
		\psfrag{l1}{\tiny$l_1$}
		\psfrag{l2}{\tiny$l_1-K_1$}
		\psfrag{l3}{\tiny$l_1+K_3$}
		\includegraphics[width=0.4\columnwidth]{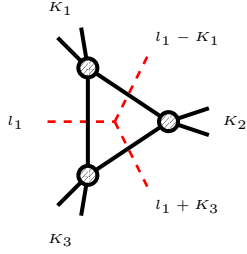}
	\end{center}
	\caption{Generic triangle cut diagram}
	\label{fig:tri}
\end{figure}

For the triple cuts we are left with a non-trivial one-dimensional integral after the on-shell
conditions have been imposed. Following the formalism of Forde \cite{Forde:intcoeffs} we can use
simple complex analysis to isolate the scalar triangle coefficients, and again reduce the problem to a purely algebraic
one. We first write down a generic triple cut integral as
\begin{align}
	T_3 &= \frac{8i\pi^3}{\pi^{D/2}}\int d^Dl_1 \prod_{i=1}^{3} \delta(l_i^2-m_i^2)A_1A_2A_3.
\end{align}
The loop momentum is chosen in a similar way to the box case but now has a free a complex parameter,
$t$:
\begin{equation}
	l_1 = a \kf 3 + b \kf 1 + t|\kf 3\ra[\kf 1| + \frac{c}{t}|\kf 1\ra[\kf 3|,
\end{equation}
where $\kf 1,\kf 3$ are defined analogously to eqs. (\ref{eq:fldefs1}-\ref{eq:fldefs3}).
The triple cut integral can then be written in terms of its pole structure in the complex $t$
plane,
\begin{align}
	T_3 = \frac{4i\pi^3}{\pi^{D/2}}\int &J_t dt 
	\sum_\sigma \bigg( \inf_t[A_1A_2A_3(l_1^\sigma)] \nonumber\\
	&+\frac{\res_{t=t_i}(A_1A_2A_3(l_1^\sigma)}{t-t_i}\bigg),
\end{align}
where the $\inf_t$ operation encodes the polynomial behaviour on the boundary of the $t$ contour.
The loop momenta, $l^\sigma$, are the complex solutions to the on-shell conditions,
\begin{equation}
	\mathcal{S} = \{l_1^2=m_1^2,(l_1-K_1)^2=m_2^2,(l_1+K_3)^2=m_3^2\}.
\end{equation}
Because the second term in the above equation has no $t$ dependence in the numerator and has at least
one additional propagator it must be associated with the scalar box coefficients which have already
been determined. Therefore the only contribution to the scalar triangle is the boundary behaviour
described by the large $t$ polynomial, $\inf_t$. For the given choice of parametrisation, the
integrals over positive and negative powers of $t$ vanish:
\begin{equation}
	\int dt J_t \,t^n = 0\quad \forall\, n\in \mathbb{Z}/\{0\}.
\end{equation}
We are then only left with the $t^0$ component which forms the complete coefficient,
\begin{equation}
	C^{m_1m_2m_3}_{4;K_1|K_2} = -\frac{1}{2}\sum_\sigma\inf_t[A_1A_2A_3(l_1^\sigma)]|_{t^0}.
	\label{eq:tricoeff}
\end{equation}

\subsection{Bubble Coefficients}

\begin{figure}[b]
	\begin{center}
		\psfrag{K1}{\tiny$K_1$}
		\psfrag{K2}{\tiny$K_3$}
		\psfrag{K3}{\tiny$K_2$}
		\psfrag{l1}{\tiny$l_1$}
		\psfrag{l2}{\tiny$l_2$}
		\psfrag{l3}{\tiny$l_3$}
		\includegraphics[width=\columnwidth]{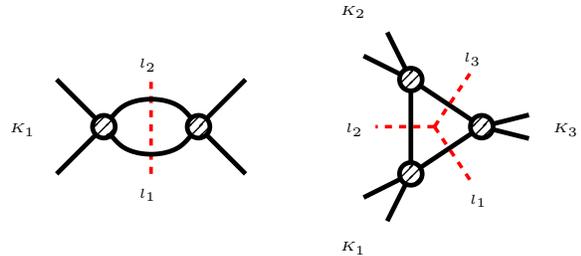}
	\end{center}
	\caption{Bubble cuts with triangle subtraction terms}
	\label{fig:bub}
\end{figure}

The bubble coefficients can be extracted from the double cut using a similar analysis to the triple
cut case, as considered in the previous section. In this case two non-trivial integrals
remain, which we choose to be parametrised by $y$ and $t$. The basis for the loop momentum in this case
requires the introduction of an arbitrary massless vector $\chi$ for which the final coefficient is
independent,
\begin{align}
	&l_1 = y \kf 1 + a(1-y) \chi + t|\kf 1\ra[\chi| \nonumber\\
	&+ \frac{b y(1-y)}{t}|\chi\ra[\kf 1|,
\end{align}
where
\begin{align}
	&\kf 1 = K_1-\frac{S_1}{\bar{\gamma}}\chi,
	&\bar{\gamma} = \AB{\chi}{K_1}{\chi}.
\end{align}
The coefficients $a$ and $b$ are solutions to the on-shell constraints,
\begin{equation}
	\{l_1^2=m_1^2,(l_1-K_1)^2=m_2^2\}.
\end{equation}
We find that there are non-vanishing integrals over both $y$ and $t$ which complicates the
procedure and requires both pure bubble and triangle subtraction terms to be included. The final
formula for the coefficient is relatively simple however,
\begin{align}
	&C_{2;K_1}^{m_1m_2} = -i\inf_t[\inf_y[A_1A_2(l_1^\sigma)]]|_{y^i\to Y_i,t^0}\nonumber\\
	&-\frac{1}{2}\sum_{\{K_2\}}\sum_{y=y_\pm}\inf_t[\res_y[A_1A_2A_3(l_1^\sigma,y_\pm)]|_{t^i\to
	T_i},
\end{align}
where $y_\pm$ are the two solutions to the additional triple cut constraint $\{(l_1+K_3)^2=m_3^2\}$. The explicit values for these non-vanishing integrals can be found in references \cite{Forde:intcoeffs,Kilgore:massuni,Badger:ttgg}, although some care must be taken when dealing with the different momentum conservation conventions.
 
\section{Application to top pair production}

\begin{figure*}[t]
	\begin{center}
		\psfrag{Q1}{\tiny$1_Q$}
		\psfrag{Q4}{\tiny$4_Q$}
		\psfrag{2}{\tiny$2$}
		\psfrag{3}{\tiny$3$}
		\psfrag{000m}{\tiny$000m$}
		\psfrag{mmm0}{\tiny$mmm0$}
		\psfrag{000}{\tiny$000$}
		\psfrag{mmm}{\tiny$mmm$}
		\psfrag{00m}{\tiny$00m$}
		\psfrag{mm0}{\tiny$mm0$}
		\psfrag{0m0}{\tiny$0m0$}
		\psfrag{m0m}{\tiny$m0m$}
		\psfrag{mm}{\tiny$mm$}
		\psfrag{00}{\tiny$00$}
		\psfrag{m0}{\tiny$m0$}
		\psfrag{0m}{\tiny$0m$}
		\includegraphics[width=9.5cm]{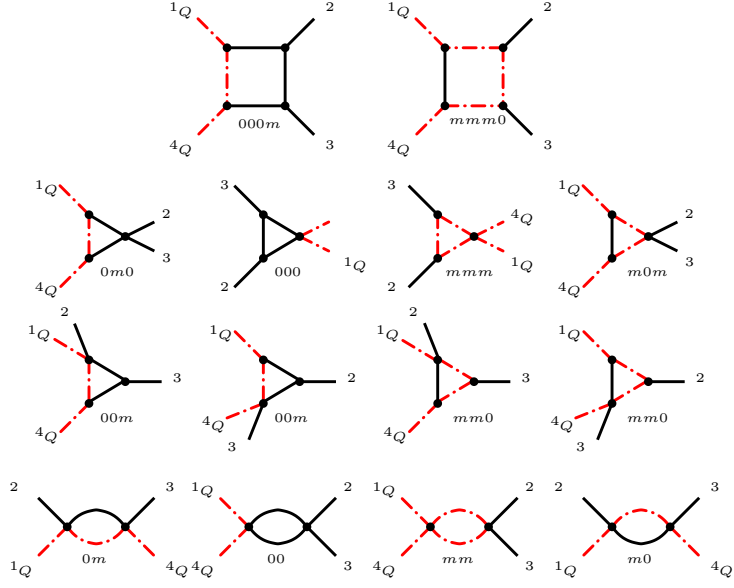}
	\end{center}
	\caption{The scalar integral basis for the $t\bar{t}gg$ amplitude. The red dot-dashed lines
	represent massive fermions whereas the solid black lines represent gluons. The internal
	mass labels appear as superscripts, i.e. $C_4^{000m}$}.
	\label{fig:ttgg_ints}
\end{figure*}

The four-point amplitudes for top pair production through gluon fusion has were computed long ago using Feynman
techniques \cite{Beenakker:ttgg} and presented analytically more recently \cite{Korner:ttgg}. As a test of our
method for extracting integral coefficients for arbitrary internal and external masses we re-compute
the cut-constructable parts of the colour ordered primitive amplitudes, $A_4^{[L]}$ and $A_4^{[R]}$, defined
by:
\begin{align}
	&A^{(1)}_4(1_Q,2,3,4_Q) = A^{[L]}_4(1_Q,2,3,4_Q)\nonumber\\&
	+\frac{1}{N_c^2}A^{[R]}_4(1_Q,2,3,4_Q)+\text{fermion/scalar loops}\nonumber\\&
	+\text{rational terms}.
\end{align}
The left-moving primitive amplitudes are those in which the fermion turns left
around the loop and the right-moving amplitudes where it turns right. The full integral basis
is shown in figure \ref{fig:ttgg_ints}. In terms of the integral basis, the cut-constructible parts
of the primitive amplitudes are:
\begin{align}
	\label{eq:Lbasis}
	&A^{[L]}_4(1_Q,2,3,4_Q) = 
	C_4^{000m}I_4^{000m}\nonumber\\&
	 +C_{3;12}^{00m}I_3^{00m}
	 +C_{3;23}^{00m}I_3^{00m}
	 +C_{3;23}^{0m0}I_3^{0m0}\nonumber\\&
	 +C_{3;23}^{000}I_3^{000}
	 +C_{2;12}^{0m}I_2^{0m}
	 +C_{2;23}^{00}I_2^{00}\nonumber\\&
	 +\frac{c_\Gamma C_1}{\epsilon}\left( \frac{\mu^2}{m^2} \right)^\epsilon\\
	\label{eq:Rbasis}
	 &A^{[R]}_4(1_Q,2,3,4_Q) = 
	 C_4^{mmm0}I_4^{mmm0}\nonumber\\&
	 +C_{3;12}^{mm0}I_3^{mm0}
	 +C_{3;23}^{mm0}I_3^{mm0}
	 +C_{3;23}^{m0m}I_3^{m0m}\nonumber\\&
	 +C_{3;23}^{mmm}I_3^{mmm}
	 +C_{2;12}^{m0}I_2^{0m}
	 +C_{2;23}^{mm}I_2^{mm}\nonumber\\&
	 +\frac{c_\Gamma C_1'}{\epsilon}\left( \frac{\mu^2}{m^2} \right)^\epsilon
\end{align}
The compact tree level input for each coefficient can be generated through on-shell recursion
relations \cite{Britto:rec}. The procedure for computing the coefficient is easily automated and has
been done both numerically and analytically. The analytic form of the final coefficient strongly
depends on the various choices for the reference momenta of both internal and external particles. In
order find the most compact representations it is convenient to begin with the most general
representation and use complex reference momenta.

Here we present the full results for the all-plus helicity configuration, the remaining helicity
amplitudes will be presented elsewhere. In this simple case the coefficients 
$C_{3;23}^{000},
C_{3;23}^{mmm},
C_{3;23}^{0m0},
C_{3;12}^{00m},
C_{3;34}^{00m},
C_{2;23}^{00}$
and
$C_{2;23}^{mm}$, all vanish leaving only 7 out of the 14 non-zero. In fact we can further simplify
the solution by making a special choice for the reference momenta $\eta_1,\eta_4$. Choosing
$\eta_1=\eta_4=p_2$ also makes the $C_4^{000m}$ coefficients vanish and the 6 non zero coefficients
are:
\begin{align}
	&C_4^{mmm0}(1_Q^+,2^+,3^+,4_Q^+) = \nonumber\\& \hspace{8mm} 
	-\frac{
im^3\la2|1|2][32]
(2\la2|1|2]+s_{23})
}{
2\la2\fl{1}\ra\la2\fl{4}\ra\la3|1|2]
}

\end{align}
\begin{align}
	&C_{3;23}^{m0m}(1_Q^+,2^+,3^+,4_Q^+) = \nonumber\\& \hspace{8mm}
	\frac{
im^3\left(s_{23}+2\la2|1|2]\right)[32]
}{
2\la2\fl{1}\ra\la2\fl{4}\ra\la3|1|2]
}
\\
	&C_{3;12}^{mm0}(1_Q^+,2^+,3^+,4_Q^+) = \nonumber\\& \hspace{8mm}
	-\frac{
im^3\left(\la2|1|2]^2-2\la2|1|3]\la3|1|2]\right)[32]
}{
2\la2\fl{1}\ra\la2\fl{4}\ra\la2|1|2]\la3|1|2]
}
\\
	&C_{3;34}^{mm0}(1_Q^+,2^+,3^+,4_Q^+) = \nonumber\\& \hspace{8mm}
	-\frac{
im^3\la2|1|2][32]
}{
2\la2\fl{1}\ra\la2\fl{4}\ra\la3|1|2]
}
\\
	&C_{2;12}^{0m}(1_Q^+,2^+,3^+,4_Q^+) = \nonumber\\& \hspace{8mm}
	\frac{
im^3\la2|1|3][23]
}{
\la2\fl{1}\ra\la2\fl{4}\ra\la2|1|2]^2}
\\
	&C_{2;12}^{m0}(1_Q^+,2^+,3^+,4_Q^+) = \nonumber\\& \hspace{8mm}
	-C_{2;12}^{0m}(1_Q^+,2^+,3^+,4_Q^+).
\end{align}
All of these expressions agree numerically with expressions extracted from the Feynman calculation of ref.
\cite{Korner:ttgg}.

\subsection{Fixing the tadpole and wave-function renormalisation coefficients}

The final step in the calculation is to fix the coefficients of the remaining $\tfrac{1}{\epsilon}$
terms coming from tadpole and wavefunction renormalisation contributions. Although it may be
possible to evaluate these terms using unitarity cuts, a number of subtleties arise, as has been
discussed in a recent numerical study \cite{Ellis:massdduni}. Instead we find it more practical
to use the universal UV and IR factorisation properties \cite{Bern:massuni,Mitov:massUV}.

Here we present a simple argument which is sufficient to find the remaining cut-constructible terms
for the $t\bar{t}gg$ amplitude. A complete analysis will appear elsewhere \cite{Badger:ttgg}. The first
step is to notice that the pure $\tfrac{1}{\epsilon}$ poles must be
proportional to the tree level amplitude, $A_4^{(0)}$. Since we know the analytic forms for the
bubble integrals,
and their coefficients this allows us to write down the coefficients of the remaining $\log(m^2)$
terms. From the basis for the primitive amplitudes, eqs.
(\ref{eq:Lbasis}-\ref{eq:Rbasis}), we see these $\log(m^2)$ terms can
be completely determined from the coefficient of the pure $\tfrac{1}{\epsilon}$ pole,
\begin{align}
	C_1 &= \alpha A_4^{(0)}-C_{2;12}^{0m}-C_{2;23}^{00}\\
	C_1' &= \alpha' A_4^{(0)}-C_{2;12}^{m0}-C_{2;23}^{mm}.
\end{align}
The coefficients $\alpha$ and $\alpha'$ are simply derived from the well known UV behaviour of the
massless amplitudes and the recent small mass factorisation of Moch and Mitov \cite{Mitov:massUV}.
The factorisation of the massive loop amplitude, $A^{(1)}(m)$, is given by,
\begin{equation}
	A^{(1)}(m) \overset{m\to0}{\to} A^{(1)}(0) + \sum_{i=1}^n Z_2^{[f(i)]} A^{(0)}(0)
\end{equation}
From the known pole structure of the massless amplitude, $A^{(1)}(0)$ , we find a contribution of $-\tfrac{3n_QC_F}{2}$
\cite{Giele:nloepem,Catani:nlosub} to the $1/\epsilon$ coefficient. It is then straightforward to
read off an additional factor of $-C_Fn_Q$ from $Z_2^{[Q]}$ for each heavy quark appearing in the amplitude \footnote{There will
also be gluon self energy corrections, $Z_2^{[g]}$, but they appear in heavy quark primitive amplitudes not
considered here.} Putting the two pieces of information together gives us:
\begin{align}
	\alpha = -\alpha' = -\frac{5}{2},
\end{align}
which matches the known result \cite{Korner:ttgg}.

\section{Conclusions}

We have shown that using the unitarity formalism of Forde \cite{Forde:intcoeffs}, and its
generalisation for arbitrary masses \cite{Kilgore:massuni}, provides an efficient method for
calculations of complicated QCD processes. We presented a simple example of a helicity amplitude
for the process of top pair production through gluon fusion while we refer the reader to
\cite{Badger:ttgg} for a complete analysis.

We foresee that future developments to include rational contributions
\cite{Giele:ddnumgu,Ossola:rational,Berger:blackhat} should
permit applications to much needed processes of top production in association with many jets in the
near future \cite{Ellis:massdduni}.

\bibliographystyle{h-elsevier2}
\bibliography{ll2008-badger}

\end{document}